\documentclass{elsarticle}
\usepackage{latexsym,amsmath,amssymb,amsbsy,graphicx}
\usepackage{numcompress}
\usepackage{lineno}
\usepackage{epstopdf}
\usepackage{color}

\DeclareGraphicsExtensions{.eps}

\def\bea{\begin{eqnarray}}
\def\eea{\end{eqnarray}}
\def\be{\begin{equation}}
\def\ee{\end{equation}}

\begin{document}

\begin{frontmatter}

\title{Radiation of fast positrons interacting with periodic  microstructure on the surface of a crystal}


\author[mymainaddress,mysecondaryaddress]{ V.\, Epp\corref{correspondingauthor}}
\cortext[correspondingauthor]{Corresponding author}
\ead{epp@tspu.edu.ru}
\address[mymainaddress]{Tomsk State Pedagogical University, ul. Kievskaya 60, 634061 Tomsk, Russia}
\address[mysecondaryaddress]{Tomsk State University, pr. Lenina 36, 634050 Tomsk, Russia}

\author[address]{J.\,G.\,Janz}
\ead{Yanc@tpu.ru}

\author[address]{V.\,V.\,Kaplin}
\ead{kaplin@tpu.ru}

\address[address]
{Tomsk Polytechnic University, pr. Lenina 34, 
634050 Tomsk, Russia}

\begin{abstract}
Radiation of  charged particles passing through a set of  equidistant ridges on the surface of a single crystal is calculated. The ridges are rectangular in shape, each  of thickness of half of the particle trajectory period at planar channeling in a thick crystal. 
Positively charged particle entering the first ridge  with angle  smaller than the critical channeling
angle is captured into channeling and changes the direction of its transversal velocity to reversed.   Between the half-wave ridges the particle moves along a straight line.
Passing through such set of half-wave crystal plates the particle  moves on quasi-undulator trajectories. Properties of the particle radiation emitted during their passage through such ``multicrystal undulator'' are calculated. The radiation spectrum in each direction is discrete, and the frequency of the first harmonic and the number of harmonics in the spectrum depends on the distance between the plates, on energy of the particles and on the averaged  potential energy of atomic planes of the crystal. The radiation is bound to a narrow cone in the direction of the average particle velocity and polarized essentially in a plane orthogonal to the atomic planes of the crystal.
\end{abstract}

\begin{keyword}
\texttt{channeling\sep radiation\sep half-wave crystal\sep relativistic particles\sep spectrum}
\end{keyword}

\end{frontmatter}

\section{Introduction}
Relativistic particles channeled in a single crystal have long been used as a source of hard electromagnetic radiation. One of the drawbacks of such a source is the limited ability of tuning the frequency of  radiation. To address this shortcoming various schemes were proposed, which combine the advantages of channeling radiation and undulator radiation. The most common schemes include periodically  bent crystals. The crystallographic planes can be bent by means of ultrasonic waves \cite{Kaplin1980, Baryshevsky1980}
 or periodic microscopic cuts of a single-crystal plate, which in this case is bent due to internal stresses \cite{Baryshevsky:2013jja, Bagli, Bellucci2003, Biryukov2004}. 
In order to get high energy radiation the crystal undulators were offered \cite{Kostyuk2013, Sushko2015341} and then realized \cite {Wistisen2014, Uggerhoj2015} with the period  much shorter than the period of particle channeling, and the bending amplitude much less than the  spacing  between the crystallographic planes.

In some cases, on the contrary, it is required to obtain more soft radiation, however, it is desirable to use a beam of sufficiently high energy particles to generate radiation of high intensity. For this purpose one can use either periodically bent crystal, or as proposed in \cite{Vorobiev1982_pat}, a set of thin crystalline plates, each changing the transversal velocity of the beam to reversal one, so that the particle trajectory in the set of the crystals looks like a zigzag line. 
Calculation of  radiation in such a device referred to as a multicrystal undulator, is the focus of this paper.

The trajectories of electrons and positrons in  a crystal plate of thickness equal to half of the particle trajectory period at planar channeling 
are investigated by numerical methods in \cite{Pivovarov_2014}. Experimental study of the passage of electrons through such a half-wave plate is described in paper \cite{Takabayashi2015}. 
The radiation emitted in a single half-wave plate was studied numerically and it was shown that the basic properties of radiation are similar to those of radiation from the arc of a circle \cite{Bagrov1983, Polozkov2015212}. Coherent superposition of the radiation fields generated in a series of half-wave plates results in the emission spectrum with specific properties that are studied theoretically in this paper.
\section{The particle trajectory}
\begin{figure}[htbp]
\center
\includegraphics[width=2.1in]{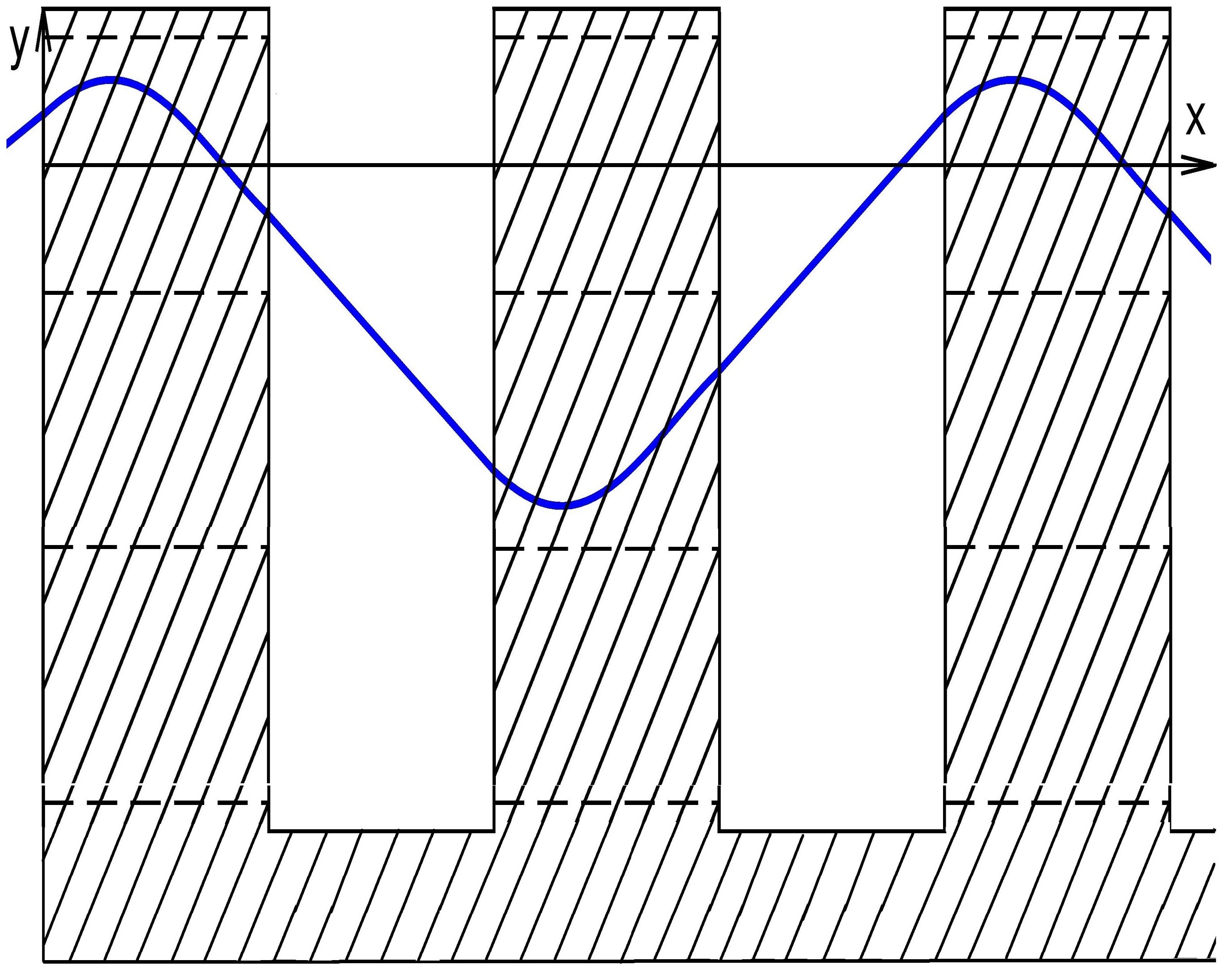}
\caption{Diagram of multicrystal undulator. The dotted lines show the crystal planes.}
\label{fig0}
\end{figure}

Consider a multicrystal undulator constructed on the surface of a single crystal as shown in Fig. \ref{fig0}.
 It consists of  a set of  equidistant ridges  rectangular in shape and each  of  half-wave  thickness.  The crystal planes which enable the channeling of positively charged particles, are orthogonal to the surfaces of the ridges. The height of the ridges is much greater than their thickness, hence, the rest of the crystal can be neglected.  Hereafter we  consider the  ridges as a set of thin crystal plates.
 Similar design was used in experiments \cite{Kaplin2000,Kaplin2001} for production of diffracted X-ray transition radiation and parametric X-ray radiation at Bragg angle, but the thickness  of the ridges was much greater than half of the channeling period.

 We choose a coordinate system in such a way that the $x$-axis is parallel to, and the $y$-axis  is orthogonal to the crystal planes. $Z$-axis  is perpendicular to the initial velocity of the  incident particle.  The thickness of each plate is $d_1$, the spacing between the plates is  $d_2$, and the distance between crystal planes equals $2a$.
We  use harmonic approximation for the average potential of the electric field between the crystal planes:
\begin{equation}\label{pot-1}
U(y)=\frac{U_0}{a^2}y^2,
\end{equation}
where $U_0$ is the depth of the potential valley.
The relativistic equations of motion of the particle in the plane $z=0$ have the form
\begin{equation}\label{eq-mot-x}
\frac{{ d}p_x}{{ d}t}=0,\quad
\frac{{ d}p_y}{{ d} t}=-\frac{2eU_0}{a^2}y.
\end{equation}
Here $ p_x$ and $p_y $ are the momentum components  
\[
p_i=\gamma m\dot x_i, \, x_i=x,y, \, \gamma=(1-\beta^2)^{1/2}, \, \beta^2=\beta_x^2+\beta_y^2,
\]
$m$ and $e$ are the mass and the charge of the particle, $\beta_i=\dot x_i/c$, the dot denotes the time derivative, $c$ is the speed of light.

Let the particle be ultrarelativistic ($\gamma\gg 1$) and the conditions of the dipole approximation are fulfilled \cite{Bordovitsyn-SR}
\begin{align}\label{dipol}
\beta_y\ll\gamma^{-1},\quad \delta\beta_x\ll \gamma^{-2},
\end{align}
where $\delta\beta_x$ is the amplitude  of $\beta_x$ variation. In this case the value of $ \gamma$ can be considered as a constant.
Integration of the equations (\ref{eq-mot-x}) in the adopted approximation gives
\begin{align}
\label{eq-motor-y2}
 x(t)=vt,\quad
y(t)=\frac{v_{0y}}{\omega_{0}}\sin{\omega_{0}t}+y_{0}\cos{\omega_{0}t},
\end{align}
where $v$ and $v_{0y}$ are the  initial particle velocity and its $y$-component, $\omega_{0}$ is the  frequency of the particle oscillations 
\begin{equation}
\omega_{0}^2=\frac{2U_0}{a^2m\gamma}.
\end{equation}

The condition of channeling is that the amplitude of  oscillations $y_0^2+(v_{0y}/\omega_0)^2$ must be less than half of the distance between the crystal planes. Hence, if a parallel beam of particles is incident upon the surface of the crystal plate, only 
particles 
with initial coordinate $y_0$ satisfying the condition
\begin{equation}\label{trajj}
y_0^2<y_m^2=a^2-\left(\frac{v_{0y}}{\omega_0}\right)^2.
\end{equation}
will be captured into the channeling mode.
The remaining particles have above-barrier transverse energy.
It follows from the above inequality that the initial transverse velocity $v_{0y}$ must satisfy  condition $|v_ {0y}|<a\omega_0$. If we introduce the angle of incidence as $\alpha=|v_{0y}|/c$, the last inequality can be written as a well known Lindhard condition
\begin{equation}
\alpha<\alpha_c=\sqrt{\frac{2U_0}{m\gamma c^2}}.
\end{equation}

The  thickness of each crystal plate is  equal to half of the trajectory period: $d_1=\pi v/\omega_{0}$. Hence, the $y$-coordinate of the particle as it leaves the first crystal plate is  $y_1=-y_0$, and the $y$-component of the velocity is   equal to $v_{y1}=-v_{y0}$. Between the first and the second crystal plates the particle moves along a stright line.
If we require that all the particles deflected by the first plate are captured into channeling by the second plate, the particle coordinate $y$ at its incident on the second plate must satisfy condition similar to (\ref{trajj}). This means that the transversal shift of the particle between the plates must make up a integer number of the distance between the crystal planes. This imposes the following condition on the distance $d_2$ between the plates
\be
d_2=2an\frac{v}{v_{0y}}=\frac{2an}{\alpha},\quad n=0,1,2 \dots\, .
\ee
Hereafter we suppose that this condition is fulfilled. Then the law of the particle motion within the second crystal plate  has the form
\begin{align}
\label{eq-motor-xy1}
 x&=vt,\\
 y&=
  -\dfrac{v_{0y}}{\omega_{0}}\sin{\omega_{0}(t-t_2)}-y_{0}\cos{\omega_{0}(t-t_2)}-2an,
 \label{eq-mot-y2}
\end{align}
where $t_2=(d_1+d_2)/v$. 
If the angle of incidence $\alpha$ is zero, then all the particles of the beam will be captured into channeling regardless of the distance $d_2$.

The first two crystal plates and two subsequent gaps  make up the spatial period of the multi-crystal undulator. The time period of motion  is equal to $T=2(d_1+d_2)/v$.

\section{Radiation}

Spectral and angular distribution of the energy radiated by the particle at quasi-periodic motion along a plane trajectory is given by  formula
\cite{Bordovitsyn-SR}

\begin{equation}\label{e59}
\frac{d{\cal E}}{ d\Omega d\omega}
=\frac{4e^2\gamma^4} {c\tilde{\omega}^2}
\frac{\sin^2\pi\nu N}{\sin^2\pi\nu}(\rho_\sigma+\rho_\pi)
|\dot{\boldsymbol\beta}(\nu)|^2,
\end{equation}
where $N$ is the number of periods of the undulator, the angular distributions of the  polarization components  $\rho_\sigma$ and $\rho_\pi$ are defined by equations
\begin{eqnarray}\label{e49ppp}
\rho_\sigma=\frac{(1-\psi^2\cos 2\varphi)^2}{(1+\psi ^2)^4}\:,\ \
\ \rho_\pi=\frac{\psi^4\sin^2 2\varphi}{(1+\psi ^2)^4},
\end{eqnarray}
and the Fourier component of acceleration is given by
\begin{eqnarray}\label{e56}
\dot{\boldsymbol\beta}(\nu)=\frac {1}{T}\int\limits_0^T \dot{\boldsymbol\beta}(t)
e^{i\tilde{\omega}\nu t}dt,\quad \nu=\frac{\omega}{2\gamma^2\tilde{\omega}}(1+\psi^2).
\end{eqnarray}
Here $\tilde{\omega}=2\pi/T$ is the frequency of the particle oscillations, $\psi=\gamma\theta$, $\theta$ is the angle between the direction of  radiation  and the channeling axis ($x$-axis), $\varphi$ is the azimuthal angle in  $yz$-plane .

\begin{figure}[htbp]
\center
\includegraphics[width=76mm]{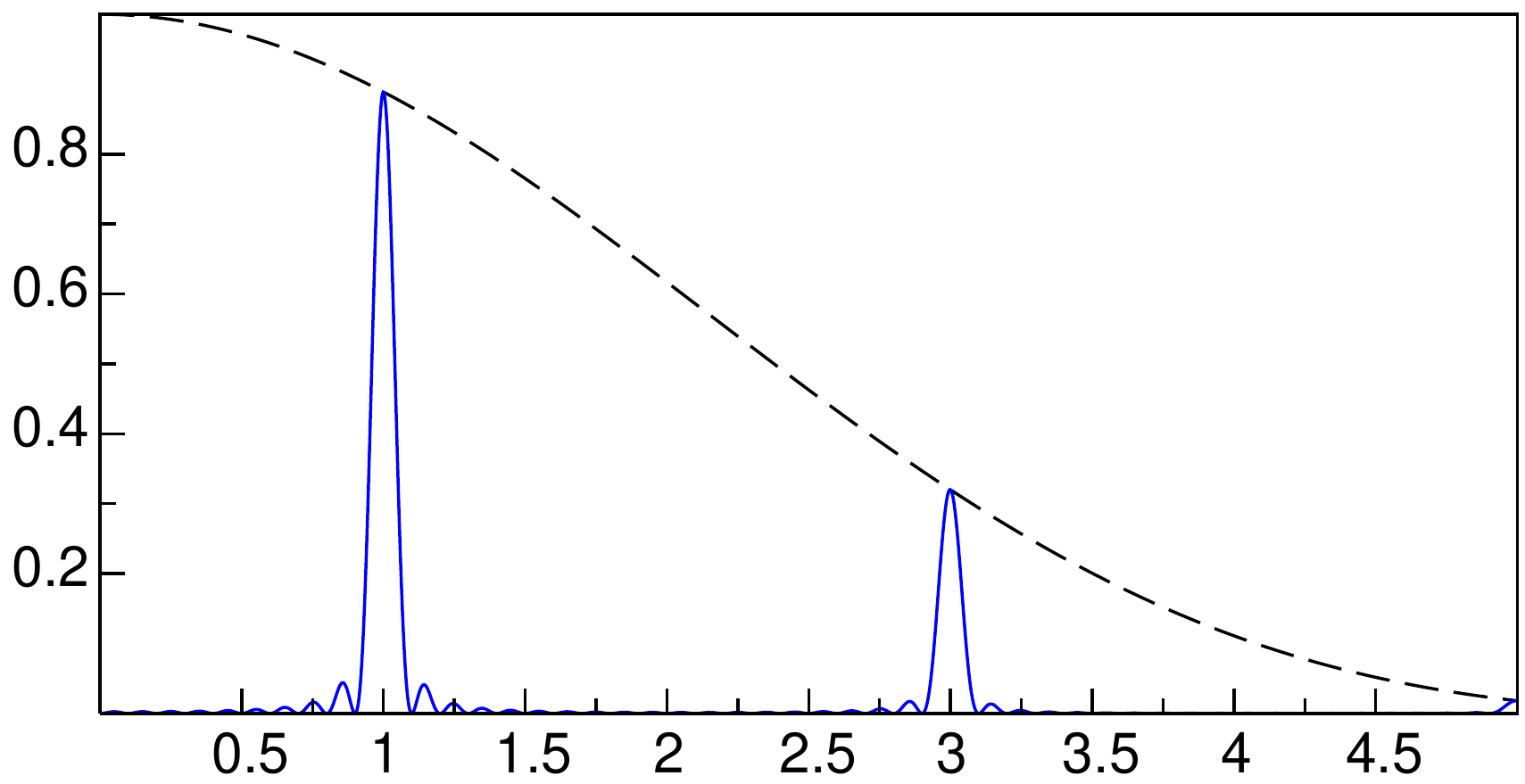}
\caption{The emission spectrum $I_1(\nu)I_2(\nu)$. $N=10, \, \eta=0.5$ }
\label{fig2}
\end{figure}

Calculating the acceleration by use of equations  (\ref{eq-motor-y2}) and (\ref{eq-mot-y2}) and substituting it in equations (\ref{e59}) and (\ref{e56}), we obtain
\begin{multline}\label{e60}
\frac{d{\cal E}}{ d\Omega d\omega}
=\frac{16e^2\gamma^4a^2\omega_0^2N^2} {\pi^2c^3}
I_1(\nu)I_2(\nu)(\rho_\sigma+\rho_\pi)\times\\
\times\left[\left(\frac{y_0\nu\eta}{a}\right)^2+\phi^2\right],
\end{multline}
where $\phi=\alpha/\alpha_c$ is the ratio of the angle of  incidence  to the critical angle of channeling, and
 $\eta=\tilde{\omega}/\omega_0<1$ is the ratio of the frequency of particle oscillations  to the  frequency of oscillation at channeling (similar to duty cycle in electronics).
The spectral functions
\begin{align}
I_1(\nu)=&\frac{\cos^2\pi\nu\eta/2}{(1-\eta^2\nu^2)^2},\nonumber\\
I_2(\nu)=&\frac{\sin^2\pi\nu N}{4N^2\cos^2(\pi\nu/2)}
\end{align}
define essential properties of radiation. In case $N\gg 1$ the function $ I_2(\nu) $ generates a narrow spectral lines of width  $\Delta \nu=N^{-1}$ in the neighborhood of odd values of $\nu$, while the function $ I_1(\nu)$ is a spectral envelope. Fig. \ref{fig2} shows the function $ I_1(\nu)$ (dashed line) and the product  $I_1(\nu)I_2(\nu)$ (solid line) for  $N=10, \, \eta=0.5$.
Function $ I_1(\nu)$ determines the width of the radiation spectrum (the number of harmonics represented in the spectrum). It follows from  definition  that  function $ I_1(\nu)$ has its  first zero where $\nu=3/\eta$. If the distance between the crystal plates   is equal to zero, i.e. if the particle is  channeling in a solid crystal, then $\tilde{\omega}=\omega_0$. In this case $\eta=1$ and only the first harmonic is presented in the emission spectrum. The number of harmonics in the spectrum increases with the reduction of $ \eta $. The frequency of the first harmonic of radiation of ultra-relativistic particles is determined by equation (\ref{e56}):
\begin{equation}
\omega_1=\frac{2\tilde{\omega}\gamma^2}{(1+\psi^2)}.
\end{equation}

Let us find the spectral density of radiation. We calculate it for crystal undulator with  large number of periods $N$.
Using the limit \cite{Bordovitsyn-SR}
\begin{equation}
\lim\limits_{N\to\infty}\frac{\sin^2\pi\nu N}{N\sin^2\pi\nu}=\sum\limits_{n=1}^\infty\delta (n-\nu)
\end{equation}
and integrating over the solid angle $d\Omega=\theta d\theta d\varphi$, we obtain the integral  spectrum of radiation 
\begin{multline}\label{e62}
\frac{d{\cal E}}{ d\omega}
=\frac{4e^2a^2\omega_0^2\gamma^2\xi N}{\pi c^3}\times\\
\times\sum\limits_{n=1}^\infty
\frac{s_n\cos^2\pi n\eta/2}{n^4(1-n^2\eta^2)^2}\left[1-(-1)^n\right]G_n\Theta(n-\xi),
\end{multline}
where $s_n=s_{n\sigma}+s_{n\pi}$ is the sum of the components of linear polarization
\be\label{harm}
s_{n\sigma}=3\xi^2-2\xi n+n^2,\quad s_{n\pi}=(\xi-n)^2,\nonumber
\ee
$\xi$ is a reduced frequency, and $G_n$ defines the influence of initial coordinate and the incidence angle on the spectrum:
\[
\xi=\frac{\omega}{2\gamma^2\eta{\omega_0}},\quad
G_n=\left(\frac{y_0 n\eta}{a}\right)^2+\phi^2,
\]
$\Theta(n-\xi)$ is the Heaviside step function.

\begin{figure}[htbp]
\center
\includegraphics[width=80mm]{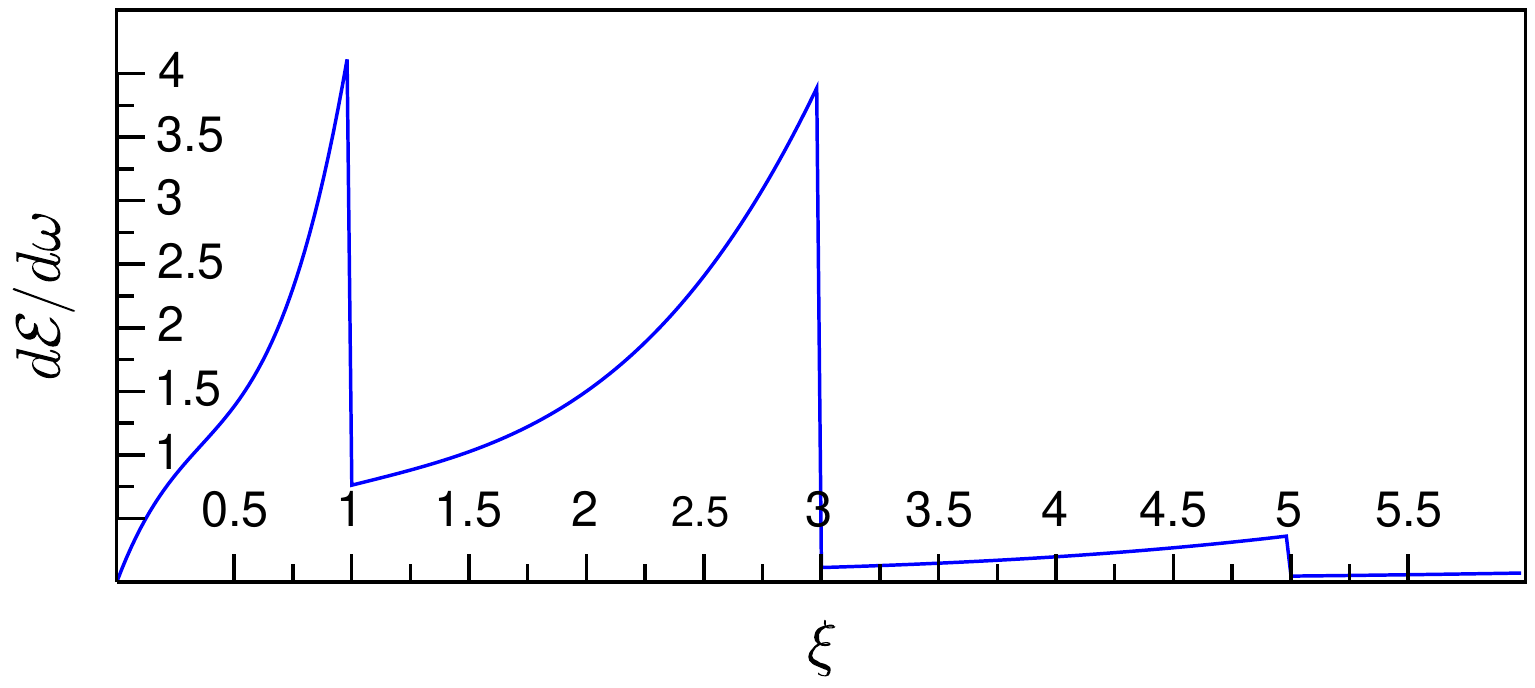}
\caption{Spectrum of radiation of parallel beam of particles. $\eta=0.5$, $\alpha=0$ }
\label{spectrum}
\end{figure}

The emission spectrum (\ref{e62}) is represented as a sum of discrete harmonics as shown in Fig, \ref{spectrum}. The shape of each harmonic 
defined by the polarization components $s_ {n\sigma}$ and $s_{n\pi}$ 
coincides with the well-known profile of undulator radiation in the dipole approximation \cite{Bordovitsyn-SR,Hofmann}.
Distribution of the radiation energy in harmonics, as can be seen from the form factor $G_n$, essentially depends on the angle of incidence of the particle in the crystal plate, its initial coordinate  $y_0$ and the distance between the plates.

\begin{figure}[htbp]
\includegraphics[width=80mm]{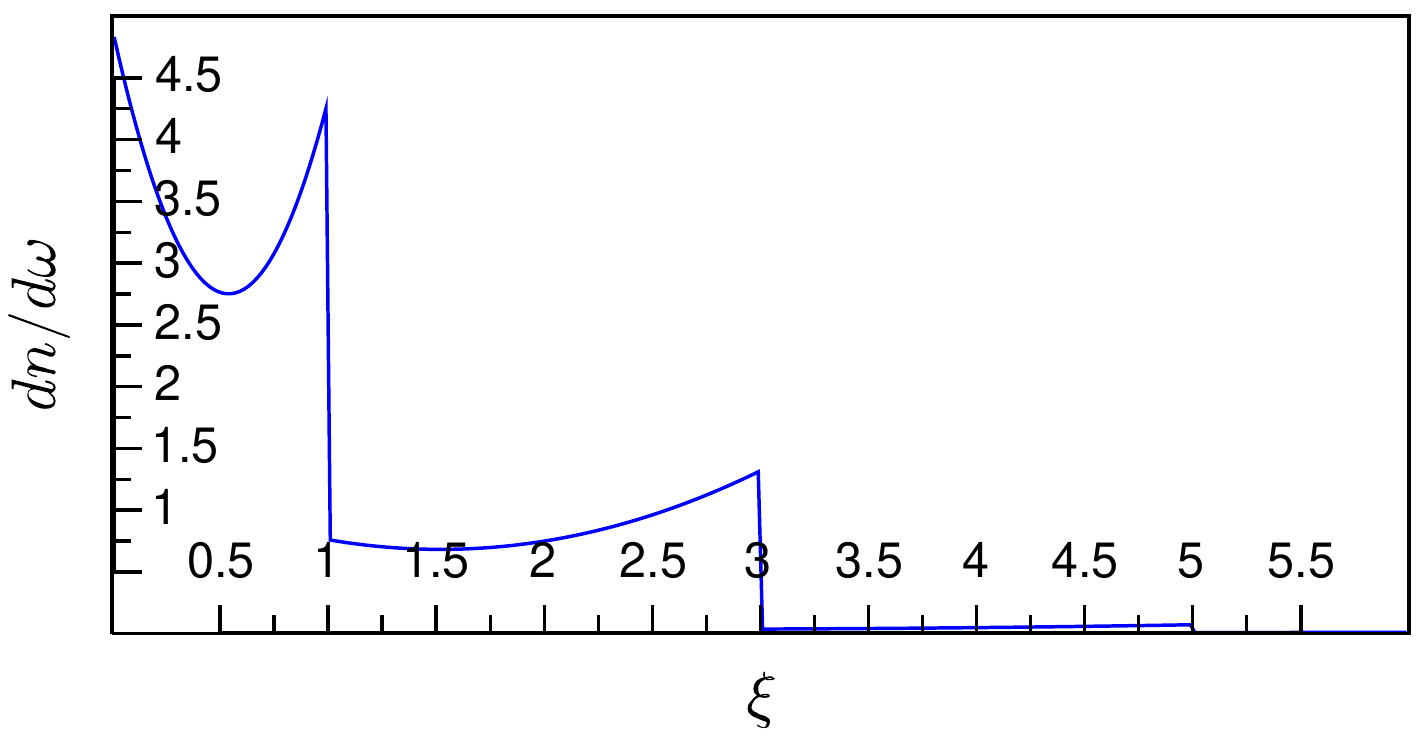}
\caption{Spectrum of radiation of parallel beam of particles for parameters $\eta=0.5$ and $\alpha=0$ }
\label{spectrum1}
\end{figure}

\begin{figure}[htbp]
\center
\includegraphics[width=80mm]{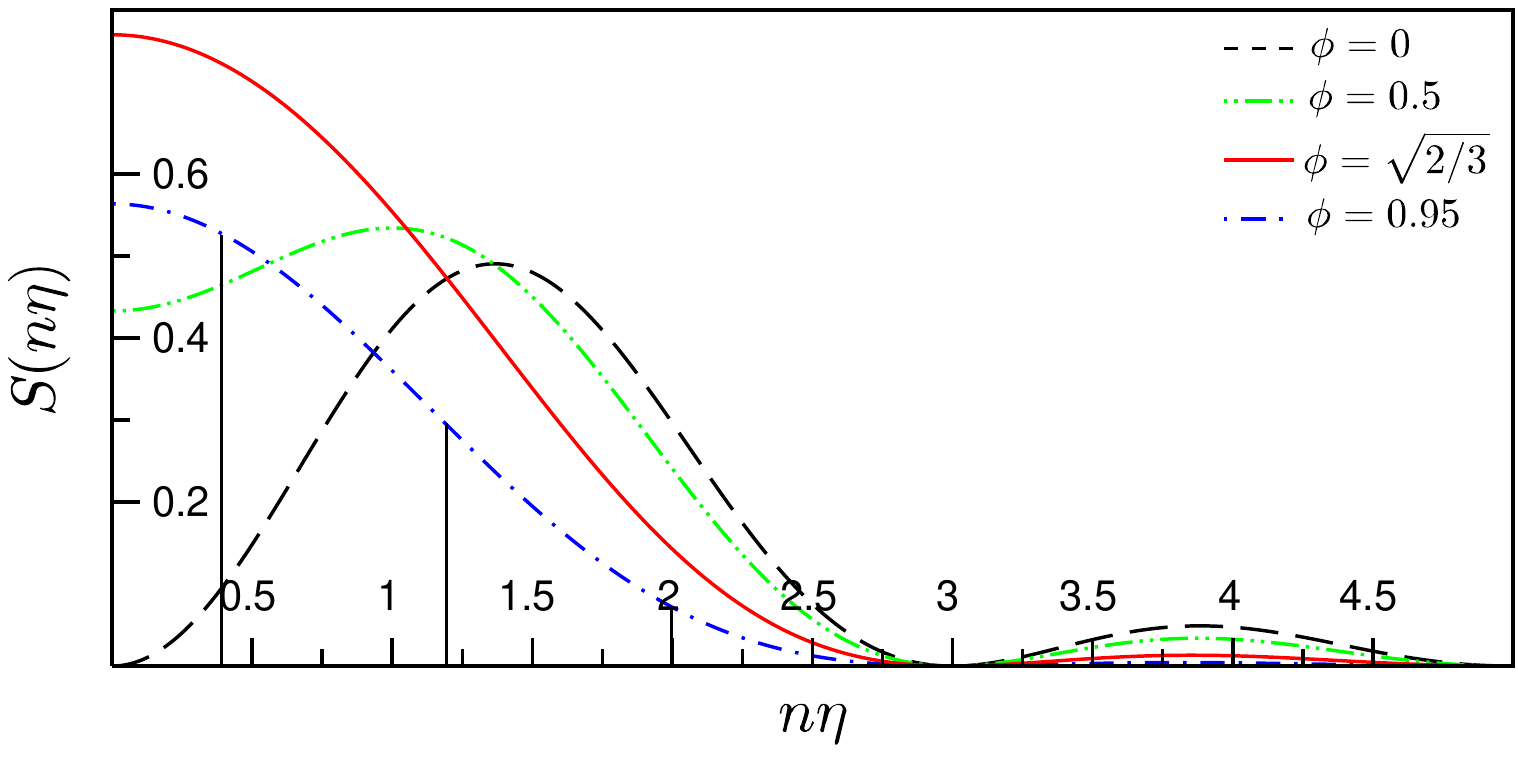}
\caption{The envelopes of spectral harmonics for different angles of  incidence $\phi$. The harmonics for undulator with $\eta=0.4$ and the angle of incidence $\phi=0.95$ are shown for example}
\label{bunch}
\end{figure}

So far we investigated the properties of radiation of a single particle. In order to find the spectral density emitted by a parallel beam of particles we average the expression (\ref{e62}) over the initial coordinate $y_0$ through the interval $2y_m$ given by equation (\ref{trajj}). Since $y_0$ is represented only in $G_n$, it will suffice to average only this multiple
\begin{equation}
{\overline G}_n=\frac{1}{2a}\int\limits_{-y_m}^{y_m}\hspace{-6pt}G_n d y_0=
(1-\phi^2)^{1/2}
\left[\frac{n^2\eta^2}{3}(1-\phi^2)+\phi^2\right]
\end{equation}
and use $\overline G_n$  instead of $G_n$ in equation (\ref{e62}). The spectrum of radiation of parallel beam of particles for parameters $\eta=0.5$ and $\alpha=0$ is plotted in Fig, \ref{spectrum}.
For the purpose of comparison with the possible experimental results the same graph is shown in Fig. \ref{spectrum1} as the number of photons  $dn$ per frequency interval $d\omega $. The spectrum consists  in this case mainly of the first and third harmonics.
The number of harmonics, making the greatest contribution to the emission spectrum is determined by the factor $\cos^2(\pi n\eta/2)/(1-n^2\eta^2)^2$, which is close to unity in the domain of  low $n\eta$ and decreases rapidly if $n\eta\gtrsim 3$. Thus, the main part of the spectrum contains about $n\sim 3/\eta$ harmonics. Furthermore, the emission spectrum depends on the angle of incidence of the particle beam on the crystal.
To explore this issue in more detail, let us find the energy of the radiation at each harmonic by integration of  the expression (\ref{e62}) over $\xi$ within a single harmonic. Taking into account the replacement of $ G_n\to{\overline G}_n$ we obtain the following
\begin{align}
&{\cal E}=\sum\limits_{n=1}^\infty({\cal E}_{n\sigma}+{\cal E}_{n\pi}),\quad {\cal E}_{n\sigma}=\frac 78{\cal E}_n,\quad {\cal E}_{n\pi}=\frac 18{\cal E}_n,\nonumber\\
&{\cal E}_n=\frac{16e^2a^2\omega_0^3\gamma^4N\eta}{3\pi c^3}S(n\eta).
\label{e611}
\end{align}
The distribution of emitted energy over the harmonics is given by the function
\begin{equation}\label{Sn}
S(n\eta)=\frac{\cos^2\pi n\eta/2}{(1-n^2\eta^2)^2}\left[1-(-1)^n\right]{\overline G}_n.
\end{equation}
This is a discrete function of odd integer values of $n$. The envelopes of this function for different angles of incidence $\phi$ are shown in Fig.\ref{bunch}
The dependence of the emission spectrum of the relative angle  $\phi$ is presented only in the factor
$\overline G_n(\phi)$.
If $\phi=0$, this factor takes the value  $\overline G_n(0)=n^2\eta^2/3$. As  the angle of incidence increases,  the value of $\overline G_n(\phi)$ increases in the low-frequency part of the spectrum ($n^2\eta^2<2$) and reaches its maximum at $\phi=\phi_m$
\begin{equation}
\overline G_n(\phi_m)=\frac{2}{3\sqrt{3-n^2\eta^2}},\quad \phi_m=\sqrt{\frac{2-n^2\eta^2}{3-n^2\eta^2}}.
\end{equation}
On further increasing of $ \phi $ this function decreases and vanishes at $\phi=1$. In the high-frequency part of the spectrum ($n^2\eta^2\ge 2$) the function $\overline G_n(\phi)$ decreases monotonically to zero with increasing of the angle of incidence.

Summing up  equation (\ref{e611}) over the harmonics  we obtain the total energy emitted by the  beam of particles, per particle
\begin{equation}\label{ful}
{\cal E}=\frac{2\pi e^2 a^2\omega_0^3\gamma^4N}{9c^3}\sqrt{1-\phi^2}\left(1+2\phi^2\right),
\end{equation}
which does not depend on $\eta$ and is distributed between the polarization components according to the ratio of 1:7. As a function of the angle of incidence, the emitted  energy is a maximum at $\phi=1/\sqrt{2}$, i.e., when the angle of incidence is $\sqrt{2}$ times less than the critical angle of channeling.
\section{Discussion}
If  a source of radiation with a frequency less than the frequency of radiation at channeling is needed, the proposed method has certain advantages compared with radiation at channeling emitted by less energetic particles. Indeed, if you want to reduce the frequency of radiation of a channeling particles $k$ times, you have to use particles with energy $k ^ {2/3}$ times less. This will cause the cone of radiation to increase $k ^ {2/3} $ times, and spectral-angular density (\ref{e60}), which is proportional to $\gamma ^ 4 $, to decrease $k ^ {8/3} $ times.  In the case of multi-crystal undulator the same decrease in frequency can be achieved if the distance between the crystal plates is equal to $d _2 = (k-1) d_1 $. Therewith the cone of radiation is not changed, and the spectral-angular density of radiation can both increase or decrease dependent on the angle of beam  incidence.

There is an interesting peculiarity of the spectral-angular density of radiation. If the distance between the crystal plates increases, then the frequency of radiation on each harmonic decreases, and the number of harmonics represented in the main part of the spectrum ($n\eta\lesssim 3$) increases in the same ratio. According to equation (\ref{ful}) the total emitted energy  does not depend on $\eta$. Hence the energy emitted at each harmonic decreases as it is  indicated by factor $\eta$ in equation (\ref{e611}). However, the spectral-angular density and the spectral density does not necessary decrease with $\eta$, because there is not an $\eta$-factor in equations  (\ref{e60}) and  (\ref{e62}). This is also seen from Fig. \ref{spectrum}: as $\eta$ decreases, the scale-factor of $\xi$-axes changes and the cut-off edge of each harmonic moves to the left in terms of $\omega$. This definitely leads to decrease of the area under the curve representing the spectrum. But the spectral density itself changes slowly according to the terms of the sum in equation (\ref{e62}).  If $\phi > 2/\sqrt {3 \pi ^ 2-20} \approx 0.645 $, then  with the decrease of $\eta$ the  spectral-angular  and the spectral density of radiation at each harmonic is growing, despite the emergence of additional harmonics.

We have considered an idealized model in order to investigate the general properties of radiation. For example, the averaged potential in the vicinity of  the crystal plane  is not harmonic. As a result, the particles channeling with high amplitude of oscillations will have a smaller spatial period and will leave the `half-wave' crystal plate at another angle. In other words, the plate being  half-wave for some particles will not be such for others. This will cause  slight scattering of the originally parallel beam of particles. Indeed, modeling of the trajectories in a more realistic potential shows that after the passage of a half-wave plate  two side maxima in the angular distribution of the beam  appear\cite{Pivovarov_2014}. This and some other factors, such as a finite number of undulator periods, the energy spread of the particles in the beam, errors in manufacturing, will lead to  broadening of the spectral lines and to  blurring of the boundaries between the harmonics shown in Fig. \ref{spectrum}. At the same time, these factors do not change significantly  the distribution of radiation energy over the harmonics.

\section*{Acknowledgments}
This work was supported by a grant of the Ministry of Education and Science of the Russian Federation under project no. 3.867.2014/K



\end{document}